\documentclass[12pt]{article}
\usepackage{epsfig,amsfonts,amssymb,amsbsy,amsbsy}
\def\cvp{\raise 2pt\hbox{,}}

\def\tr{\mathop{\rm tr}\nolimits}
\def\im{\mathop{\rm Im}\nolimits}

\def\diag{\mathop{\rm diag}\nolimits}
\def\plb#1#2#3{{\it Phys.\ Lett.\ }{\bf B #1} (#2) #3}
\def\npb#1#2#3{{\it Nucl.\ Phys.\ }{\bf B #1} (#2) #3}
\def\prl#1#2#3{{\it Phys.\ Rev.\ Lett.\ }{\bf #1} (#2) #3}
\def\jhep#1#2#3{{\it JHEP\ }{\bf #1} (#2) #3}
\def\prd#1#2#3{{\it Phys.\ Rev.\ }{\bf D #1} (#2) #3}
\def\atmp#1#2#3{{\it Adv.\ Theor.\ Math.\ Phys.\ }{\bf #1} (#2) #3}
\def\cmp#1#2#3{{\it Comm.\ Math.\ Phys.\ }{\bf #1} (#2) #3}
\def\suN{{\rm SU}(N)}
\def\seff{S_{\rm eff}}
\def\rc{r_{\rm c}}
\begin{document}
%
%
\pagestyle{empty}
{\parskip 0in
\hfill PUPT-1990

\hfill LPTENS-01/09

\hfill hep-th/0106192}

\vfill
\begin{center}
{\large\bf The large $N$ limit of ${\cal N}=2$ super Yang-Mills,}\\
\medskip
{\large\bf fractional instantons and infrared divergences}

\vspace{0.4in}

Frank F{\scshape errari}\footnote{On leave of absence from Centre 
National de la Recherche Scientifique, Laboratoire de Physique 
Th\'eorique de l'\'Ecole Normale Sup\'erieure, Paris, France.}\\
\medskip
{\it Joseph Henry Laboratories\\
Princeton University, Princeton, New Jersey 08544, USA}\\
\smallskip
{\tt fferrari@feynman.princeton.edu}
\end{center}

\vfill\noindent
We investigate the large $N$ limit of pure ${\cal N}=2$
supersymmetric gauge theory with gauge group ${\rm SU}(N)$ by
using the exact low energy effective action. 
Typical one-complex dimensional sections of the moduli space
parametrized by a global complex mass scale $v$ display three 
qualitatively different regions depending on the ratio between $|v|$
and the dynamically generated scale $\Lambda$.
At large $|v|/\Lambda$, instantons are exponentially
suppressed as $N\rightarrow\infty$.  
When $|v|\sim\Lambda$, singularities due to massless dyons occur.
They are densely distributed in rings of calculable thicknesses in the
$v$-plane. At small $|v|/\Lambda$, instantons disintegrate into fractional 
instantons of charge $1/(2N)$. These fractional instantons give non-trivial 
contributions to all orders of $1/N$, unlike a planar diagrams 
expansion which generates a series in $1/N^{2}$, implying the presence 
of open strings. We have explicitly calculated the fractional 
instantons series in two representative examples, including
the $1/N$ and $1/N^{2}$ corrections.
Our most interesting finding is that the $1/N$ 
expansion breaks down at singularities on the moduli space due to severe 
infrared divergencies, a fact that has remarkable consequences.
\vfill
\begin{flushleft}
June 2001
\end{flushleft}
\newpage\pagestyle{plain}
\baselineskip 16pt
%
\section{Introduction}
Quantum ${\rm SU}(N)$ gauge theory has a single free parameter, the number 
of colours $N$, or equivalently the number of interacting gluons $N^{2}-1$.
Our best chance to understand the strongly 
coupled dynamics of gluons is probably to perform an analysis of the theory at 
large $N$ \cite{tHooft}. In the present work we will use exact 
results available for theories with eight supercharges (${\cal N}=2$ 
supersymmetry) \cite{SW,Oz} in order to try to better understand the 
large $N$ limit of four dimensional gauge theories.
Some basic qualitative features were first discovered 
by the author while working on simple two dimensional toy models, which 
can be supersymmetric \cite{fer1}, but also non supersymmetric 
\cite{fer2,fer3}. In four dimensions, we can unfortunately provide 
explicit calculations for ${\cal N}=2$ theories only.

The paper is 
organized as follows. In Section 2, we give a general qualitative 
discussion of some of our main results, without entering into technical 
details. In Sections 3 and 4, we discuss in turn the structure of the moduli 
space at large $N$, the nature and properties of the 
large $N$ expansion of the low energy observables,
and two full large $N$ calculations of representative 
Seiberg-Witten period integrals.
In Section 5 we discuss some open problems.
\section{General discussion}

{\it Large $N$ in ${\it SU}(N)$ gauge theories and planar diagrams}

\noindent
From the point of view of perturbation theory, the 
large $N$ expansion is a reordering of Feynman diagrams with respect to 
their topology, with diagrams of genus $h$ being proportional to $1/N^{2h}$. 
The original perturbative series has zero radius of 
convergence, but the large $N$ reordering can produce convergent series in 
the gauge coupling constant at each order in $1/N^{2}$, thus providing 
a non-perturbative treatment. The series in $1/N^{2}$ is reminiscent 
of the genus expansion of an oriented closed string theory \cite{pol} of 
coupling constant $\kappa\sim 1/N$, and one may hope that a full 
non-perturbative equivalence between string theories and gauge theories 
could exist. This is supported by the empirical facts that hadrons are 
found on a Regge trajectory, and that quarks seem to be confined by the 
collimation of Faraday flux lines in string-like structures.
In a supersymmetric context, one can make this idea more 
precise by using the fact that D-branes, on which gauge theories live,
can be viewed as solitons in closed string theories \cite{malda}.
\vfill\eject
\noindent {\it Large $N$ and instantons}

\noindent
A general amplitude in gauge theory can pick up contributions of different
types. In addition to the terms having a Feynman diagram representation
that are discussed in the context of the 't Hooft
expansion, non-trivial field configurations invisible in Feynman
diagrams can in principle contribute to the path integral.
The most popular configurations of this type are instantons \cite{SP, H2}.
Instantons are responsible for important semi-classical effects like
tunneling. In the case of ${\cal N}=2$ supersymmetric gauge theories, on
which we will focus in the present paper, a non-renormalization theorem
\cite{NRth} implies that  
the low energy effective action $\seff$ up to two derivatives 
terms picks up only a trivial one-loop term from Feynman diagrams, 
but has an infinite series of instanton contributions. From the point of
view of the large $N$ limit, however, instantons are exponentially 
suppressed, 
and thus do not seem to play an interesting r\^ole \cite{Witins}. The 
instanton action is indeed proportional to $N$ in the 't Hooft scaling
$g_{\rm YM}^{2}\propto 1/N$. For
${\cal N}=2$ super Yang-Mills, this seems to imply that $\seff$ at large
$N$ is simply given by the one-loop formula. One would have to look at
higher derivative terms in order to obtain a non-trivial large $N$
expansion.

More precisely, and as was emphasized in \cite{Witins}, the effects of
instantons of size $1/v$ are proportional in the one-loop approximation to
\begin{equation}
\label{instcont}
e^{-8 \pi^2 /g_{\rm YM}^2} = \left( {\Lambda\over v}\right)^{\beta_0}\cvp
\end{equation}
where $g_{\rm YM}$ is the gauge coupling constant at scale $v$, $\Lambda$ the
dynamically generated scale of the theory, and $\beta_0\propto N$ is given
by the one-loop $\beta$ function. The one-loop formula (\ref{instcont}) is
exact for ${\cal N}=2$ super Yang-Mills, with $\beta_0 = 2N$. It suggests
that the only smooth limit of instanton contributions when
$N\rightarrow\infty$ is zero \cite{Witins}. Large instantons (small $v$),
if relevant, would produce catastrophic exponentially large contributions,
and if one is willing to assume that the large $N$ limit makes sense
the only physically sensible conclusion is that instantons
are irrelevant variables. In real-world QCD, instantons of all sizes can
potentially contribute, and this led Witten to argue that the instanton gas
must vanish \cite{Witins}. In Higgs theories, like ${\cal N}=2$ super
Yang-Mills, the Higgs vevs introduce a natural cutoff $v$ on the size of
instantons, and for $v$ large enough (``weak coupling'') the instanton gas
can exist but is just negligible at large $N$. At small $v$ (``strong
coupling''), we run into
the same difficulties as in QCD, and instantons must somehow disappear. 

\noindent {\it Large $N$ and fractional instantons}

\noindent
An interesting puzzle is to try to understand the nature of the physics at
strong coupling in the large $N$ limit. 
Unlike in non-supersymmetric theories where one may argue
that the physics is dominated by planar diagrams, in
${\cal N}=2$ gauge theories we have already pointed out that
some important observables do not have any non-trivial contribution from
Feynman diagrams. In their derivation of the low energy effective action
$\seff$, Seiberg and Witten \cite{SW} made use of the 
electric/magnetic duality transformations of the low energy abelian theory.
These transformations are very useful to understand the physics near 
singularities where magnetically charged particles become massless, but 
they say nothing about the large $N$ limit at strong
coupling. The purpose of the present paper is to elucidate this problem.
We will see that instantons actually do not literally vanish, 
but disintegrate into fractional instantons of topological
charge $1/(2N)$ (or a multiple of $1/(2N)$).
Such fractional instantons give contributions of order
\begin{equation}
\label{fracinc}
\left( {\Lambda\over v}\right)^{\beta_0/2N}={\Lambda\over v}\,\cvp
\end{equation}
and can obviously survive at large $N$. The fact that fractional instantons
could play a r\^ole in gauge theories has been suspected for some time,
independently of the large $N$ approximation.
In the case of ${\cal N}=1$ supersymmetric gauge theories,
chiral symmetry is broken by a gluino condensate
$\langle\lambda\lambda\rangle = \Lambda^3$. Since $\beta_0 = 3N$ in this
context, we see that the gluino condensate can be interpreted as coming
from a fractional instanton of charge $1/N$ (a recent discussion of the
gluino condensate can be found in \cite{hol}; see also \cite{dorey} for a
discussion of the closely related problem of ${\cal N}=1$ superpotentials
in confining vacua). In the following we will see that 
fractional instantons do play a fundamental
r\^ole in ${\cal N}=2$ gauge theories as well, when the theory is analysed 
at large $N$.

Fractional instantons is a new type of contribution in the large $N$ limit. 
Though it is known that Feynman diagrams 
generate a series in $1/N^{2}$ and that instantons must be exponentially 
suppressed, it is not obvious what is the analogous statement for 
fractional instantons. We will explicitly demonstrate below that in ${\cal 
N}=2$ gauge theories, fractional instantons generate a series in 
$1/N$. The leading contribution can be of the same order of magnitude as 
the one coming from planar graphs. The first correction is then of order 
$1/N$, and would dominate any subleading $1/N^{2}$
corrections from Feynman diagrams. 
This proves directly from the field theory point of 
view that the string theory dual
must contain open strings in addition to the familiar closed strings. This 
fact is actually consistent with our present knowledge on string 
duals to ${\cal N}=2$ gauge theories. In the supergravity 
approximation, the closed string background dual to such theories usually 
has unphysical naked singularities called ``repulsons'' \cite{repulson}. 
String theory does actually make sense on such backgrounds thanks to the 
``enhan\c con'' mechanism \cite{enhancon}: the constituent branes expand and 
form a finite shell excising the singularity. The open strings 
responsible for the fractional instantons contributions must be open 
strings attached to these branes. Such open strings, 
and the fact that they would produce unusual $1/N$ corrections on the field 
theory side, do not seem to have been discussed in the literature. It 
would be extremely interesting if they could be used to match with
non-trivial results of the type we are going to derive on the gauge theory 
side. 

\noindent {\it Large $N$ and singularities}

\noindent
Possibly the most interesting result derived in this paper
concerns the behaviour 
of the large $N$ expansion at singularities on the moduli space. These 
results were anticipated in our study of two dimensional models 
\cite{fer1,fer2,fer3}.

On the moduli space of vacua, the gauge group $\suN$ is
broken down to ${\rm U}(1)^{N-1}$, and generically the low energy theory 
is a simple pure ${\cal N}=2$ abelian gauge theory. For some 
particular, ``critical,'' values of the moduli, however, some additional 
hypermultiplets can become massless \cite{SW}. In the pure $\suN$ case on 
which we focus, such critical points can only be observed at strong 
coupling (the W bosons have then a mass of order $\Lambda$) and correspond 
to massless magnetically charged particles called dyons. Since those couple 
non-locally to the original photons, it is a priori not obvious what the 
low energy theory can be, but we know from \cite{SW} that 
the dyons couple locally to dual photons and generally produce a trivial free 
abelian theory in the IR. When the light particles are not mutually local 
with respect to each other, which can be achieved by adjusting several 
moduli to special values (higher critical point), a genuinely non-trivial 
CFT develops in the IR \cite{AD}. In gauge theories with quark 
hypermultiplets, very general types of critical behaviour can be 
obtained, see e.~g.\ \cite{ADE}.

Commonly, trying to find a good approximation scheme to describe a 
non-trivial critical point can be subtle in field theory.
A typical example is 
$\phi^{4}$ theory in dimension $D$. The theory has two parameters, the 
bare mass $m$ (or ``temperature'') and the bare coupling constant $g$. By 
adjusting the temperature, we can go to a point where we have massless 
degrees of freedom, and a non-trivial Ising CFT in the IR.
The difficulty is that the renormalized fixed point coupling $g_{*}$
is large, and thus ordinary perturbation theory in $g$ fails. It is 
meaningless to try to calculate universal quantities like critical 
exponents as power series in $g$, since those are $g$-independent. 
Either the tree-level, $g$-independent contributions are exact and the 
corrections vanish (this occurs above the critical dimension, which is
$D_{\rm c}=4$ for the Ising model, and we have a trivial
fixed point $g_{*}=0$ well described by 
mean field theory), or the expansion parameter corresponds to a relevant 
operator and corrections to mean field theory are plagued by untamable IR 
divergencies. The way out of this 
problem is to use expansion parameters on which the low energy CFT depends 
(corresponding in some sense to marginal or nearly marginal operators): we 
can use an $\epsilon$ expansion by going to $4-\epsilon$ dimensions, or a 
large $N$ expansion by considering ${\rm O}(N)$ vector models and ${\rm O}(N)$ invariant 
critical points.

The critical points on the moduli space of ${\cal N}=2$ 
super Yang-Mills are very similar to the Ising critical point below the
critical dimension,
and $1/N$ is very similar to the $\phi^{4}$ coupling $g$. On any 
generic finite dimensional submanifold of the moduli space,
one finds critical 
points, characterized by a set of critical exponents \cite{AD,ASW,ADE},
that are independent of the number of colours $N$ of the gauge 
theory in which the CFT is embedded. These critical exponents cannot 
consistently be calculated in a $1/N$ expansion.
Even the simple monopole critical points that are known to be trivial
are not described consistently by 
the $N\rightarrow\infty$ limit of the original gauge theory,
because electric-magnetic duality is not 
implemented naturally in this approximation scheme. 
We will indeed explicitly demonstrate in 
Section 4 that, though the leading large $N$ 
approximation for the critical points is smooth (``mean field theory''),
the $1/N$ corrections are IR divergent.

Note that by adjusting a large number $\sim N$ of moduli, one can also 
study $N$-dependent critical points. For these, the large $N$ expansion can 
certainly be consistent. The $N$ special vacua where a maximum number $N-1$ of 
mutually local dyons become massless and which where studied in \cite{DS} 
because of their relationship with ${\cal N}=1$ vacua are of this type.
\section{The structure of moduli space at large $N$}
\subsection{Brief review of ${\cal N}=2$ super Yang-Mills}
The fields of pure $\suN$, ${\cal N}=2$ supersymmetric gauge theory all
transform in the adjoint representation of the gauge group and 
make up an ${\cal N}=2$ vector multiplet which contains the gluons, 
gluinos and a complex scalar field $\Phi$ with scalar potential 
\begin{equation}
\label{scalarpot}
V = {1\over g^2}\, \tr \lbrack\Phi ,\Phi^\dagger\rbrack^2 .
\end{equation}
The $D$-flatness conditions $\lbrack\Phi ,\Phi^\dagger\rbrack =0$ are
solved by
\begin{equation}
\label{Dflat}
\langle\Phi\rangle = \diag (a_1 ,\ldots ,a_N)
\end{equation}
with
\begin{equation}
\label{tran}
\sum_{i=1}^N a_i =0.
\end{equation}
Classically, the gauge group is generically broken to ${\rm U}(1)^{N-1}$, 
with W bosons of masses $m_{ij} = \sqrt{2} |a_i - a_j|$. When some of the
$a_i$s coincide, the gauge symmetry is enhanced.
Quantum mechanically, the moduli space is parametrized by the vevs of the
gauge invariant operators
\begin{equation}
\label{ui}
u_i = \langle\tr \Phi^i\rangle ,\quad 2\leq i \leq N ,
\end{equation}
or equivalently by the $\phi_i$s defined implicitly by the relations
\begin{equation}
\label{phis}
u_k = \sum_{i=1}^N \phi_i^k\ , \qquad \sum_{i=1}^N \phi_i =0.
\end{equation}
The $a_i$s are non-trivial functions of the $\phi_i$s,
with $a_i\simeq \phi_i$ only at weak
coupling. In particular, it turns out to be impossible to
have $a_i = a_j$, which implies that
a non-abelian gauge group is never restored in the quantum theory.

The low energy effective action is then generically a pure abelian ${\rm
U}(1)^{N-1}$ theory. The leading terms in a derivative expansion 
can be written in ${\cal N}=1$ superspace as
\begin{equation}
\label{seff}
S_{\rm eff} = {1\over 4\pi}\, \im\int\! d^4 x \left[ \int\! d^4\theta\,
\partial_i {\cal F}(A) \bar A^i + {1\over 2}\,\int\! d^2\theta\,
\partial_i\partial_j {\cal F}(A) W^i W^j \right],
\end{equation}
where the $(A_i,W_i)$s are the massless abelian
${\cal N}=2$ vector multiplets satisfying $\sum_i A_i =0$
and ${\cal F}(A)$ is a holomorphic function called the prepotential.  

According to Gauss's law, the central charge $Z$ 
of the supersymmetry algebra, and the masses 
\begin{equation}
\label{BPSmass}
M_{\rm BPS} = \sqrt{2}\, |Z|
\end{equation}
of the BPS states, only depend on the values of the fields at large 
distances, and thus they can be expressed
in terms of $\cal F$ only. Introducing the dual variables
\begin{equation}
\label{addef}
a_{Di} = {\partial {\cal F}\over\partial a_i}
\end{equation}
we have \cite{SW}
\begin{equation}
\label{Z}
Z = \sum_{i=1}^N \bigl( q_i a_i + h_i a_{Di} \bigr) 
\end{equation}
where the integers $q_i$, $\sum_i q_i =0$, and $h_i$, $\sum_i h_i=0$,
are the electric and magnetic quantum numbers respectively. 
Low energy electric/magnetic duality is manifest 
on the formulas (\ref{seff}) 
and (\ref{Z}) which are ${\rm Sp}(2(N-1),{\mathbb Z})$ invariant, and  
it turns out that the variables $a$ and $a_D$ are most naturally 
interpreted as sections of an ${\rm Sp}(2(N-1),{\mathbb Z})$ bundle.
\vfill\eject
If one introduces the genus $N-1$ hyperelliptic curve $\cal C$ defined by
\begin{equation}
\label{curve}
{\cal C}: \quad Y^2 = \prod_{i=1}^N \bigl(X-\phi_i\bigr)^2 - 
\Lambda^{2N} = P(X)^2 - \Lambda^{2N} 
\end{equation}
equiped with a canonical basis of homology cycles $(\alpha_i ,\beta_i)$,  
and the differential form
\begin{equation}
\label{SWd}
\lambda_{\rm SW} = {X\, dP\over 2i\pi Y}\cvp
\end{equation}
then $a$ and $a_D$ are simply given by \cite{SW,Oz}
\begin{equation}
\label{period}
a_i = \oint_{\beta_i}\lambda_{\rm SW} ,\quad a_{Di} = \oint_{\alpha_i}
\lambda_{\rm SW}.
\end{equation}
It is natural to rewrite the equation for $\cal C$ as
\begin{equation}
\label{curve2}
Y^2 = \bigl( P(X) - \Lambda^N \bigr) \bigl( P(X) + \Lambda^N \bigr) =
P_+(X) P_-(X) =
\prod_{i=1}^N \bigl( X-X_{i,+} \bigr) \bigl( X-X_{i,-} \bigr)
\end{equation}
and to view the curve $\cal C$ as two copies of the complex plane with
cuts joining the branching points $X_{i,+}$ and $X_{i,-}$. 
The ``electric'' contour $\beta_i$ defining $a_i$ 
encircles the cut from $X_{i,+}$ to $X_{i,-}$. At weak coupling 
(small $\Lambda$), one then recovers $a_i\simeq\phi_i$. For our purposes,
however, it is not particularly useful to try to make a distinction between
electric and magnetic variables. At strong coupling, the two notions are 
mixed by the non-trivial monodromies \cite{SW}.
The important point is that there is a one-to-one correspondence between
possible BPS states and the integer homology of the curve $\cal C$, 
in such a way that to any cycle $\gamma\in H_1({\cal C},{\mathbb Z})$ 
we can associate the central charge
\begin{equation}
\label{Z2}
Z(\gamma) = \oint_\gamma \lambda_{\rm SW} .
\end{equation}
Singularities on the moduli space are obtained when $H_1({\cal C},{\mathbb
Z})$ degenerates, or equivalently when two of the branching points
coincide. Since the polynomials $P_+$ and $P_-$ obviously have no common
roots, this can happen only when the discriminants of $P_+$ or $P_-$ 
independently vanish,
\begin{equation}
\label{singloc}
\Delta(P_+) = \prod_{i<j} (X_{i,+} - X_{j,+})^2 = 0\quad {\rm or}\quad
\Delta(P_-)=\prod_{i<j} (X_{i,-} - X_{j,-})^2 = 0.
\end{equation}
\vfill\eject
\subsection{The large $N$ limit}
\subsubsection{Generalities}
At large $N$, it is convenient to parametrize the $N-1$ dimensional
moduli space by the density function
\begin{equation}
\label{rhodef}
\rho_{N}(\phi) = {1\over N} \sum_{i=1}^N \delta^{(2)} (\phi - \phi_i)
\end{equation}
which satisfies
\begin{equation}
\label{rhoc}
\int\rho_{N} (\phi)\, d^2\phi = 1\ ,\quad \int \phi\rho_{N}(\phi)\, d^2\phi =0.
\end{equation}
At weak coupling, $\rho_{N}(\phi)$ simply gives the density of complex
eigenvalues of the Higgs field. The differences $\phi_i-\phi_j$ are related
to the masses of the W bosons. If the lightest W has a mass of order $m_{\rm
W}$, then the heaviest W will generically have a mass of order $M_{\rm
W}\sim \sqrt{N} m_{\rm W}$. Particles having a mass growing with $N$ are 
problematic in the large $N$ expansion. This is due to the fact that the 
amplitude for the propagation of a state of mass $M$ for a time $t$
involves the factor $\exp (-iMt)$ (this argument was used long ago by 
Witten for baryons \cite{witbar}). A consistent large $N$ limit, for which 
propagators of W bosons have a smooth expansion, can thus 
be achieved only if the heaviest Ws have a mass of order $N^{0}$, 
and thus the lightest Ws have generically a mass of order  
$1/\sqrt{N}$. For one dimensional distributions 
corresponding to the case where all the $\phi_{i}$s are aligned, the 
lightest Ws have a mass of order $1/N$. This correct large $N$
scaling was already considered in \cite{DS}. Under the above conditions, 
the distribution $\rho_{N}(\phi)$ will typically have an 
$N\rightarrow\infty$ limit $\rho (\phi)$ which is the 
sum of a smooth function with compact support plus $\delta$ 
function terms. We will consider examples below.

A good strategy to study the large $N$, $N-1$ dimensional, moduli space, is 
to consider well-chosen finite dimensional subspaces. The simplest 
subspaces $\cal M_{\rho}$
are one-complex dimensional and are parametrized by a complex mass 
scale $v$. One picks up a particular fixed distribution $\phi_{i}$ 
characterized by $\rho(\phi)$ and considers $\phi_{i}^{(v)} = v \phi_{i}$ 
or equivalently
\begin{equation}
\label{rhov}
\rho_{v}(\phi) = {1\over v^{2}}\, \rho\Bigl( {\phi\over v} \Bigr)\cdotp
\end{equation}
When $\rho(\phi)$ is a smooth function, then the limit $v\rightarrow\infty$ 
is a weak coupling limit since the lightest Ws have a 
mass proportional to $v$. If $\rho(\phi)$ has some $\delta$ function 
singularities, which means that a large number (of order $N$) of the 
$\phi_{i}$s coincide, then we are at strong coupling for any $v$.
The study of the simple subspaces ${\cal M}_{\rho}$ can reveal 
a great deal of physics, including the transition from weak coupling to 
strong coupling and the presence of the simplest singularities. To obtain 
more complicated singularities ($k$th order critical points), 
one has to consider $k$ dimensional subspaces, or introduce quark flavors 
and quark mass parameters. For the 
purposes of the present paper, it is enough to stick to the
one-dimensional subspaces ${\cal M}_{\rho}$ defined by (\ref{rhov}).
It is then natural to introduce the dimensionless ratio
\begin{equation}
\label{rdef}
r = {v\over\Lambda}
\end{equation}
and the polynomials
\begin{equation}
\label{poldef}
p(x) =\prod_{i=1}^{N} \bigl( x-\phi_{i} \bigr)\, ,\qquad
p_{\pm}(x) = p(x) \mp 1/r^{N},
\end{equation}
and to rewrite the fundamental equations (\ref{curve}), (\ref{SWd}) and 
(\ref{Z2}) in terms of dimensionless variables,
\begin{equation}
\label{curve3}
{\cal C}:\quad y^{2} = p(x)^{2} - 1/r^{2N} = p_{+}(x)\, p_{-}(x),
\end{equation}
\begin{equation}
\label{SWd2}
\lambda =\lambda_{\rm SW} \bigm/ v = x\, dp \bigm/ \bigl( 2 i \pi y \bigr) ,
\end{equation}
\begin{equation}
\label{z}
z(\gamma) = Z(\gamma) \bigm/ v = \oint_{\gamma}\lambda .
\end{equation}
\subsubsection{Singularities}
The locus of singularities $\cal S$ corresponds to the points
where the discriminants (\ref{singloc}) vanish, or 
equivalently when one of the polynomials $p_{+}$ or $p_{-}$ has a multiple 
root. The discriminants are themselves polynomials of degree $N-1$ in 
$r^{N}$, and thus we will have generically $2N(N-1)$ singularities where a dyon 
becomes massless on ${\cal M}_{\rho}$. These singularities are arranged 
on $2(N-1)$ concentric circles.
To better understand the structure of $\cal 
S$, it is useful to study first the large $N$ distribution of the 
roots of the polynomials $p_{\pm}$. The equation $p_{\pm}(x)=0$
can be rewritten 
\begin{equation}
\label{rpp}
\ln \left| r^{N}p(x) \right| + i\arg \left( \pm r^{N}p(x) \right) = 0.
\end{equation}
The real part of the equation is of order $N$ while the imaginary part is 
of order $N^{0}$. In the leading large $N$ approximation, both $p_{+}=0$ 
and $p_{-}=0$ thus reduce to
\begin{equation}
\label{rpp2}
E_{N}(x)\buildrel\hbox{\footnotesize\rm def}\over = |p(x)|^{1/N} =  1/|r|.
\end{equation}
The $N\rightarrow\infty$ limit of the ``envelope'' $E_{N}$ is easily 
evaluated,
\begin{equation}
\label{env}
E_{\infty}(x) = \exp\int\! d^{2}\phi\, \rho(\phi) \ln |\phi -x|.
\end{equation}
When $x$ is in the support of $\rho$, the above approximate
formula only gives an average value of $E_{N}$, which in fact
has very sharp wells
around the $\phi_{i}$s, $E_{N}(\phi_{i}) = 0$. In the large $N$ expansion, 
the wells are approximated by infinitely thin lines joining the smooth 
envelope $E_{\infty}$ (see Figure 1 for an example). 
Any correction to this picture would be exponentially small at 
large $N$, an instanton effect. At large $|x|$,
\begin{equation}
\label{envas}
E_{\infty}(x) = |x| \, \left(1 + {\cal O}\bigl(1/|x|\bigr)\right).
\end{equation}
Equation (\ref{rpp2}) is now easily solved.
When $1/|r| < \min_{x\in{\mathbb C}} 
E_{\infty}(x)$, the plane $z=1/|r|$ can only intersect 
$z=E_{N}$ along the thin wells: the roots of $p_{+}$ and $p_-$
coincide with the
classical roots $\phi_{i}$. This corresponds to the instanton dominated 
region, and those are suppressed at large $N$. 
When $|r|$ decreases and
$1/|r| > \min_{x\in{\mathbb C}} E_{\infty}(x)$, $z=1/|r|$ and $z=E_{N}$ 
continues to meet along some of the thin wells, but the roots nearest 
to the minimum of $E_{\infty}$ are now arranged along an inflating curve. 
This curve gives the general shape of the ``enhan\c con'' discussed in 
\cite{enhancon}. Note that the enhan\c con exists for all $r$
if the distribution $\rho$ has some $\delta$ function singularities. 
It has in general several connected components associated with the
different connected components of the support of $\rho$. 
At large $1/|r|$, all the roots are located on the 
enhan\c con whose various connected components eventually merge to form 
asymptotically a circle of radius $1/|r|$ according to (\ref{envas}).

\begin{figure}
\centerline{\epsfig{file=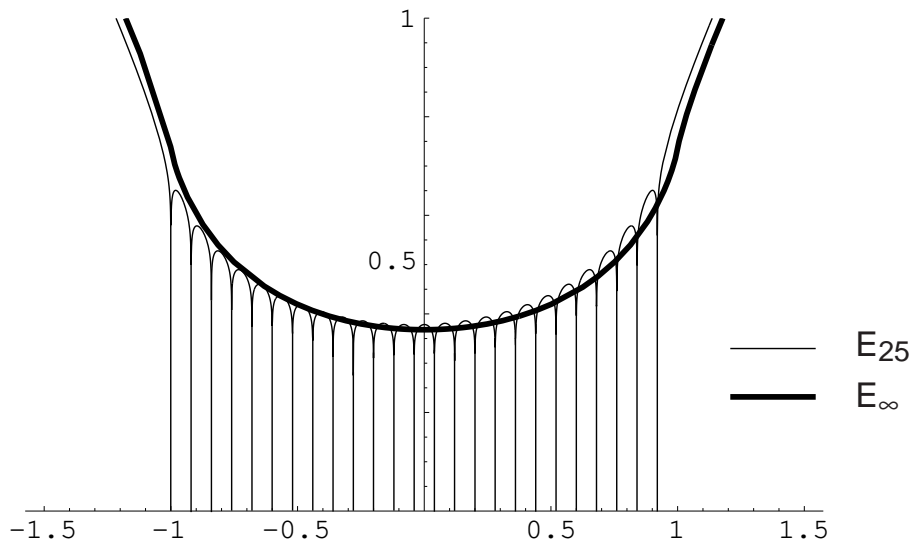,width=12cm}}
\caption{Plot of the function $E_{N=25}(x)$ 
corresponding to the distribution (\ref{unidis1}) with
$\phi_{i} = 2(i-1)/N -1$ ($1\leq i\leq N$) (thin line)
together with the envelope $E_{\infty}$ (thick line).
\label{envfig}}
\end{figure}

For example, let us consider the simple uniform distribution
\begin{equation}
\label{unidis}
\rho_{0}(\phi) = \left\{ \matrix{ 1/\pi\ {\rm if}\ |\phi|<1\hfill\cr
                              0\ {\rm if}\ |\phi|>1 .\hfill\cr} \right.
\end{equation}
It yields an envelope
\begin{equation}
\label{envuni}
E_{\infty,0}(x)= \left\{ \matrix{ e^{(|x|^{2}-1)/2}\ {\rm if}\ |x|<1\hfill\cr
              |x|\ {\rm if}\ |x| >1.\hfill\cr} \right.
\end{equation}
For $|r|>\sqrt{e}$, all the roots are classical. At $|r|=\sqrt{e}$, the 
enhan\c con begins to inflate. It is a circle of radius $\sqrt{1-\ln 
|r|^{2}}$. At $|r|=1$, the enhan\c con has eaten up all the classical 
roots, and for $|r|<1$ it is simply a circle of radius $1/|r|$. For the 
purpose of illustration, we have depicted in Figure 1 the case of a 
unidimensional uniform distribution
\begin{equation}
\label{unidis1}
\rho_{1}(\phi) = \left\{ \matrix{ 
                          \delta(\im (\phi))/2\ {\rm if}\ |\phi|<1\hfill\cr
                              0\ {\rm if}\ |\phi|>1 \hfill\cr} \right.
\end{equation}
for which
\begin{equation}
\label{envuni2}
E_{\infty,1}(x\in {\mathbb R}) = \left\{ \matrix{
  (1-x)^{(1-x)/2} (-1-x)^{(1+x)/2} /e \quad {\rm if}\ x<-1\hfill\cr
  (1+x)^{(1+x)/2} (1-x)^{(1-x)/2} /e  \quad {\rm if}\ -1<x<1.\hfill\cr
  (1+x)^{(1+x)/2} (-1+x)^{(1-x)/2} /e \quad {\rm if}\ x>1.\hfill\cr} \right.
\end{equation}

Now that we understand the location of the roots of $p_{\pm}$, it is 
straightforward to deduce the singularity locus. Generically, roots are 
either equal to the classical values, or are smoothly distributed along the 
enhan\c con. Singularities can then occur only when the classical roots are  
eaten up by the inflating enhan\c con, or when two connected components of 
the enhan\c con merge.
This is particularly transparent on Figure 1, where real 
roots join the enhan\c con by pairs and then become complex. We thus 
deduce that the singularities are densely distributed on rings in the 
$r$-plane corresponding to the melting of classical roots with the enhan\c 
con (singularities of the first class) 
or with the melting of several connected components of the enhan\c con 
with each other (singularities of the second class).
In Figure 2 we have represented the 
space ${\cal M}_{\rho_{0}}$ with the associated singularity locus $\cal S$
(only singularities of the first class are present in this example).
We will see in Section 
4 that the large $N$ expansion is afflicted with  power-like divergences 
near singularities of the second class, while logarithmic divergencies 
also mix up near singularities of the first class.

\begin{figure}
\centerline{\epsfig{file=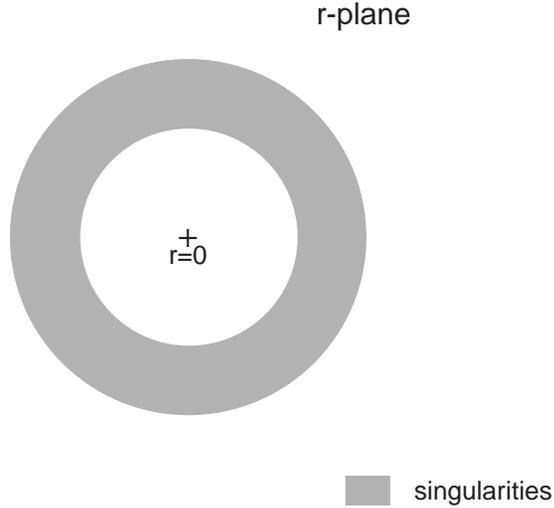,width=10cm}}
\caption{The one-complex dimensional subspace 
${\cal M}_{\rho_{0}}$ corresponding to the uniform distribution 
(\ref{unidis}) with the associated ring of singularity $\cal S$.
\label{singfig}}
\end{figure}
\subsubsection{Special cases}
We have discussed above how the singularity locus $\cal S$
can be deduced for a
given density $\rho$. It is amusing to study the inverse question:
can we find a density $\rho$ that would correspond to an a priori
given locus $\cal S$? This amounts to inversing the functional relation
(\ref{env}) between $E_{\infty}$ and $\rho$. If we limit ourselves to the
consideration of unidimensional distributions (by constraining
all the $\phi_i$s to be real), the mathematical problem is very similar to
the one encountered in the study of the leading $N\rightarrow\infty$ 
limit of simple matrix integrals \cite{matmod}. If we introduce
\begin{equation}
\label{omega}
\omega (x) = \int\! d\phi\, {\rho (\phi)\over x-\phi }\cvp
\end{equation}
we have
\begin{equation}
\label{invenv1}
\rho(\phi) = {i\over 2\pi}\, \Bigl( \omega(\phi + i\epsilon) - \omega(\phi
- i\epsilon) \Bigr),
\end{equation}
and (\ref{env}) yields 
\begin{equation}
\label{invenv}
{1\over E_\infty }\, {dE_\infty\over dx} = {1\over 2}\,
\Bigl( \omega(\phi + i\epsilon) + \omega(\phi - i\epsilon) \Bigr).
\end{equation}
As an application, let us try to find a simple unidimensional smooth
density $\rho$ with a connected compact support $[-1,1]$ for which
all the singularities are distributed on a single circle (a singularity
ring of zero thickness). From the discussion of Section 3.2.2
it is clear that this can only happen when
$dE_\infty /dx =0$ on the support of $\rho$.
The only solution to (\ref{invenv}) consistent with
$\omega(x)\sim 1/x$ at infinity is then
\begin{equation}
\label{omesol}
\omega (x) = {1\over\sqrt{(x-1)(x+1)}}\cvp
\end{equation}
and (\ref{invenv1}) implies
\begin{equation}
\label{chedis}
\rho(\phi) = {1\over \pi\sqrt{1-\phi^2}}\cdotp
\end{equation}
This is the density distribution of the roots of the
Chebyshev polynomials $T_N$, for which $\phi_i = \cos 
\bigl( \pi (1/2 + i)/N\bigr)$. This very special distribution was studied in
\cite{DS}. Equation (\ref{env}) then gives the radius of the singularity
circle $|r|=2$. Clearly, many generalisations could be studied.
\subsubsection{Corrections to $N=\infty$}
The singularity rings of the first class
separate an outer purely instanton-dominated region from an 
inner strongly coupled region which cannot be described by semi-classical 
physics (the outer region only exists when $\rho$ does not have any 
$\delta$ function singularities). In the outer region, corrections at 
large $N$ are exponentially suppressed. To the contrary, we can a priori
expect non-trivial corrections in the inner, 
strongly coupled, regions of moduli space. We will elucidate the nature 
and the main properties of these corrections in Section 4, but we can have 
a first instructive look at them by investigating the envelope $E_{\infty}$
or equivalently the shape of the enhan\c con on a simple example.
Let us choose $\phi_{i} = 2(i-1)/N -1$ as in Figure 1, and let us calculate
the corrections to (\ref{envuni2}) for $x>1$. By applying Euler's formula
\begin{equation}
\label{euler}
{b-a\over n}\, \sum_{k=0}^{n-1} f\Bigl( a + k {b-a\over n} \Bigr) =
\int_{a}^{b}\! f - {\bigl( b-a \bigr) \bigl( f(b)-f(a) \bigr) \over 2n} +
{\cal O}(1/n^{2})
\end{equation}
we obtain
\begin{equation}
\label{nc1}
E_{N,1}(x) = E_{\infty ,1}(x) \Bigl( 1 - {1\over 2 N}\, 
 \ln {x-1\over x+1}  + {\cal O}(1/N^{2}) \Bigr) \quad {\rm for}\ x>1 .
\end{equation}
This formula displays two of the main features of the corrections to 
$N=\infty$: they are of order $1/N$ ({\it not} of order $1/N^{2}$) and 
they diverge near the singularity locus which corresponds here to 
$|r|\rightarrow (1/2)e^{+}$ or equivalently $x\rightarrow 1^{+}$ 
(see (\ref{rpp2}) and (\ref{envuni2})).
A third feature, that the corrections are given by series of
fractional instantons, is also to be expected if one consider that 
(\ref{rpp2}) depends on $1/r$.

\section{Calculation of the $1/N$ and $1/N^2$ corrections}

This Section is devoted to a detailed study of the corrections to the
leading $N\rightarrow\infty$ approximation in two particular, but generic,
examples, encompassing the cases of both types of singularities (first 
class and second class).
Since we want to consider the purely
strongly coupled region, where the physics cannot be described by
instantons, we can put most of the singularities at infinity. Yet,
we will keep one singular circle at finite distance, so that we can 
discuss the physics associated with singularities.

\subsection{An example with a first class singularity}

We choose the distribution $\rho_N$ to be
\begin{equation}
\label{sperho}
\rho_N(\phi) = {N-1\over N}\, \delta \bigl(\phi + 1/N\bigr) + 
{1\over N}\,\delta\bigl(\phi + 1/N -1\bigr).
\end{equation}
The associated function $E_N$ and envelope $E_\infty$ are depicted in
Figure 3. The parameter $r$ will be real and positive, without loss of 
generality (the final formulas can be trivially analytically continued). We 
have a singularity of the first class
corresponding to the melting of the classical root 
$x_{\rm cl} = 1 - 1/N$ with the enhan\c con.

\begin{figure}
\centerline{\epsfig{file=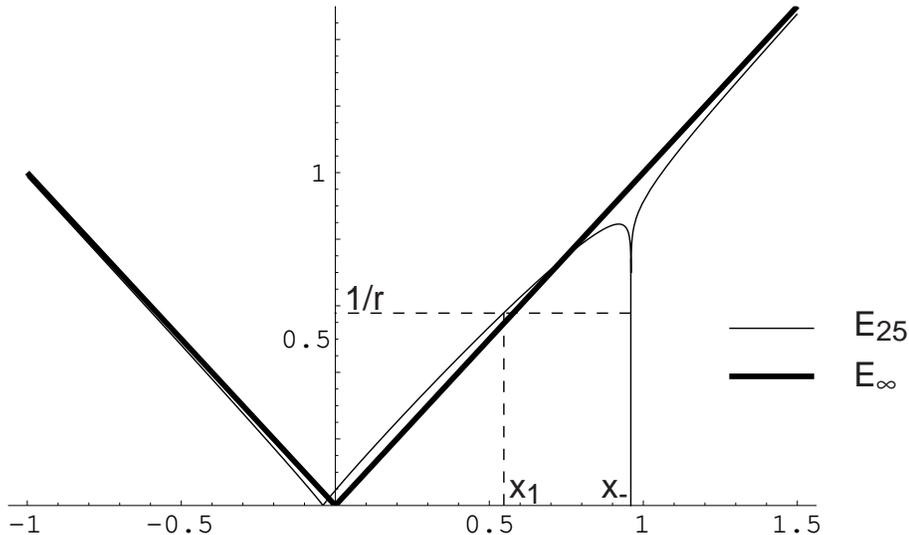,width=12cm}}
\caption{Plot of the function $E_{N=25}(x)$ corresponding to the 
distribution (\ref{sperho}) (thin line) together with the envelope
$E_{\infty}$ (thick line). The roots $x_{-}$ and $x_{1}$ of the polynomial 
$p_{-}$ discussed in the text (equations (\ref{xm}) and (\ref{x1})) are also 
indicated for a particular value of $1/r$.
\label{fig3}}
\end{figure}
\subsubsection{Infrared divergences and fractional instantons}
A very simple, yet very instructive, calculation one can do is 
the critical
value $\rc$ of $r$ for which the singularity occurs. The leading
$N\rightarrow\infty$ approximation $\rc =1$
can be read off immediadely from Figure 3. The exact value can be found by
noting that it corresponds to the point where the two positive real roots  
of the polynomial $p_-$ defined by (\ref{poldef}) coincide,
\begin{equation}
\label{rcrit}
\rc = {N\over (N-1)^{1-1/N}}\cdotp
\end{equation}
The large $N$ expansion of the exact formula (\ref{rcrit}) has strange
looking logarithmic terms,
\begin{equation}
\label{rcritapp}
\rc = 1 + {\ln N + 1\over N} + o(1/N).
\end{equation}
A standard, well-behaved, large $N$ expansion cannot produce such $(\ln N)
/N$ corrections. Remarkably, the same kind
of $(\ln N)/N$ corrections to the critical
parameter were found recently by the author in a study of a two dimensional
QFT that was argued to be very similar to gauge theories 
with Higgs fields \cite{fer2,fer3}. These $(\ln N)/N$ corrections were the
consequence of the breakdown of the large $N$ expansion near the critical
point due to infrared divergences (a full calculation can be found in
Appendix C of \cite{fer3}). We will see below that the qualitative
physics in four dimensions is strictly similar to the physics previously
discussed in two dimensions \cite{fer1,fer2,fer3}.

The next simple calculation one can obviously do is the large $N$
expansion of the roots of the polynomials $p_+$ and $p_-$. These roots also
enter explicitly in the Seiberg-Witten period integrals (\ref{z}). We will
focus on the two positive real roots of the polynomial $p_-$ which exists
for $r>\rc$ and coincide at $r=\rc$. One of these two roots that we call
$x_-$ is simply
\begin{equation}
\label{xm}
x_- =1-1/N - 1/r^N + {\cal O}\bigl( 1/r^{2N} \bigr).
\end{equation}
The non-trivial corrections to the classical root $x_{\rm cl}= 1-1/N$
are exponentially small. The other root $x_1$ has a non-trivial and very
interesting large $N$ expansion that can be straightforwardly deduced
from (\ref{rpp}), 
\begin{equation}
\label{x1}
x_1 = {1\over r} - {r + \ln (r-1)\over N r} + {1\over 2N^2 r}\,
\Bigl( \left( \ln (r-1) \right) ^2 - {2 r \ln(r-1)\over r-1} \Bigr)
+ {\cal O}\bigl(1/N^3 \bigr).
\end{equation}
This very simple formula displays all the features that were advertised in 
Section 2. In particular, by writing $\ln (r-1) = \ln r + \ln (1-1/r) = 
\ln r -\sum_{k\geq 1}1/(kr^{k})$, we see that each order in $1/N$ is given 
by a series of fractional instantons of fractional charge $1/(2N)$, mixed 
with the one-loop diagram contribution in $\ln r$. At $r=1$, the 
corrections are blowing up like the logarithm of the mass of the lightest 
degrees of freedom, an 
infrared divergence. Using (\ref{xm}) and (\ref{x1}) to solve 
$x_{-}=x_{1}$, one can recover the expansion (\ref{rcritapp}) with the 
$(\ln N)/N$ correction.

Of course, roots like $x_{1}$ are not directly observables. The physics is 
encoded in the Seiberg-Witten periods (\ref{z}), and though they depend on 
the roots, one might argue that cancellations could occur and that the 
final result could be smooth. It is also conceivable that only 
corrections in $1/N^{2}$ could show up in the final results. To answer
rigorously these questions, we will compute in the next subsection 
the Seiberg-Witten period integral $z$ corresponding to the cycle $\gamma$ 
which vanishes at $r=\rc$. It turns out that the only potential divergences 
come from (\ref{x1}) in this case, 
and that the qualitative features of the expansion 
for $x_{1}$ and for $z$ are the same.
\subsubsection{The calculation of the period}
We thus proceed to compute the large $N$ expansion of
\begin{equation}
\label{SWint}
z = \oint_{\gamma}\lambda = {1\over i\pi}\, \int_{x_{1}}^{x_{-}}\! dx\, 
{xp'(x)\over\sqrt{p(x)^{2}-1/r^{2N}}}\cvp
\end{equation}
with
\begin{equation}
\label{pexpl}
p(x) = (x + 1/N)^{N-1} (x-1+1/N).
\end{equation}
It is convenient to trade the variable $x$ for $u=1/(r(x+1/N))$, in terms 
of which
\begin{equation}
\label{SWint2}
z = {N\over i\pi r}\, \int_{u_{-}}^{u_{1}} {du\over u^{2}}\,
{\bigl(1-r u/N\bigr)\bigl(1-(1-1/N)r u\bigr)\over
\sqrt{(1-ru)^{2} - u^{2N}}}\cvp
\end{equation}
with
\begin{eqnarray}
\label{uvar}
u_{-}&=&{1\over r}\, \Bigl( 1 + 1/ r^{N} + {\cal O}(1/r^{2N})\Bigr),\\
u_{1}&=& 1 + {\ln(r-1)\over N} + {1\over 2N^{2}}\, \Bigl(
\left( \ln(r-1) \right)^{2} + {2r\over r-1}\, \ln(r-1) \Bigr) + 
{\cal O}(1/N^{3}).\\ \nonumber
\end{eqnarray}
At large $N$, it is tempting to neglect the term $u^{2N}$ in 
(\ref{SWint2}), which is 
exponentially small on most of the integration region. This is correct 
except for $u\simeq u_{1}\simeq 1$, when $u^{2N}$ is not negligible,
and for $u\simeq u_{-}\simeq 1/r$, when $(ru -1)^{2}$ can be smaller than 
$u^{2N}$. We will thus distinguish three different contributions to $z$,
\begin{equation}
\label{zsplit}
z = z_{0} + z_{<} + z_{>}\, ,
\end{equation}
with
\begin{eqnarray}
\hskip -.5cm z_{0} \!\!\! &=& \!\!\! {N\over i\pi r}\! 
\int_{u_{-}}^{u_{1}}\! {du\over u^{2}}
\Bigl(1-{r u\over N}\Bigr)\Bigl(1-{(N-1)r u\over N}\Bigr)
{1\over ru-1}\cvp\label{z0}\\
z_{>}\!\!\! &=&\!\!\! {N\over i\pi r}\! 
\int_{U}^{u_{1}}\! {du\over u^{2}}
\Bigl(1-{r u\over N}\Bigr)\Bigl(1-{(N-1)r u\over N}\Bigr)\!
\left( {1\over\sqrt{(r u -1)^{2} - u^{2N}}} - {1\over r u -1} \right)
\!\!\cvp\label{zp}\\ 
z_{<}\!\!\! &=&\!\!\! {N\over i\pi r}\! 
\int_{u_{-}}^{U}\! {du\over u^{2}}
\Bigl(1-{r u\over N}\Bigr)\Bigl(1-{(N-1)r u\over N}\Bigr)\!
\left( {1\over\sqrt{(r u -1)^{2} - u^{2N}}} - {1\over r u -1} \right)
\!\!\cvp\label{zm}\\
\nonumber
\end{eqnarray}
where $U\in ]u_{-},u_{1}[$ is an arbitrary constant, for example $U=(u_-
+ u_1)/2$.
The integral $z_0$ can be explicitly evaluated, and its expansion is
given by
\begin{equation}
\label{z0f}
{i\pi z_0\over N} = -{r-1\over r} + \Bigl( 1 - {1\over N} \Bigr)\, \ln r
+ {r-1\over Nr}\, \ln (r-1) + {\bigl( \ln(r-1) \bigr)^2\over 2N^2 r} +
{\cal O}\bigl( 1/N^3 \bigr).
\end{equation}
Finding the asymptotic expansion of $z_>$ at large $N$ is a bit more
subtle. The integral would be exponentially small if not for the
integration region near $u_1$. The term $u^{2N}$ cannot be neglected 
in a region of size $1/N$ around $u_1$. The idea is to use a new
variable $w$ defined by
\begin{equation}
\label{nvar}
u = u_1 - w/N.
\end{equation}
The large $N$ expansion of
the integrand can then be straightforwardly deduced, 
and one is left with several definite integrals on the
variable $w\in [0, N(u_1 -U)]$. The integration region can
be extended to
$[0,\infty [$ at the expense of neglecting exponentially suppressed terms.
The calculation of the $1/N$ and $1/N^2$ corrections involve five
non-trivial definite integrals that can be explicitly evaluated.  
Instead of writing down all these complicated formulas, we have
illustrated the procedure on a simpler, but qualitatively similar, example
in the Appendix. Fortunately, the final result is simple-looking,
\begin{equation}
\label{zpf}
{i\pi z_>\over N} = -{\ln 2\over N r} + {1\over N^2 r}\,
\left( \Bigl( \ln(r-1) + r + {1\over 2}\ln 2 \Bigr)\ln 2 -
{\pi^2\over 24} \right).
\end{equation}
The evaluation of $z_<$ proceeds along the same lines, except that now the
relevant integration region around $u=u_-$ is exponentially small. The
asymptotic expansion to all orders in $1/N$ can then be obtained in this
case by changing the variable to $w$ defined by $u = u_- + w/r^{N+1}$. We
get
\begin{equation}
\label{xmf}
{i\pi z_<\over N} = {N-1\over N^2}\, \ln 2 + {\cal O}\bigl( 1/r^N \bigr).
\end{equation}
Putting all the contributions together, we finally end up with
\begin{eqnarray}
\label{zfinal}
&&\hskip -1cm
{i\pi z\over N} = -{r-1\over r} + \ln r + {1\over N}\, \left(
-\ln r + {r-1\over r}\, \ln 2 + {(r-1)\ln(r-1)\over r}\right) \\
\nonumber
&& \hskip 1cm
+{1\over N^2 r}\, \left( {1\over 2}\, \bigl( \ln(r-1) \bigr)^2 + 
(\ln 2)\ln(r-1) + {1\over 2}\, (\ln 2)^2 - {\pi^2\over 24} \right)
+ {\cal O}\bigl( 1/N^3 \bigr).\\ \nonumber
\end{eqnarray}
This final formula displays all the remarkable properties of the large $N$
expansion of ${\cal N}=2$, $\suN$ super Yang-Mills theory in four
dimensions that we have already discussed at lenght in the previous
Sections. We note that genuine divergencies when $r\rightarrow 1$
show up only at order $1/N^{2}$, but non-analytic terms are already
present at order $1/N$.

\subsection{An example with a second class singularity}

Let us now pick $N$ to be even for convenience and consider
the case of the distribution
\begin{equation}
\label{newdis}
\rho_{N}(\phi) = {1\over 2}\, \Bigl( \delta (\phi -1 ) + \delta (\phi + 1) 
\Bigr)
\end{equation}
which corresponds to the polynomial $p(x)= (x^{2}-1)^{N/2}$ (see 
(\ref{poldef})).
The enhan\c con has two components at large $|r|$ which merge at a second 
class singularity for $|r|=1$. The shape of the enhan\c con,
together with the roots of 
$p(x)^{2} - 1/r^{2N}$ are depicted in Figure 4.

\begin{figure}
\centerline{\epsfig{file=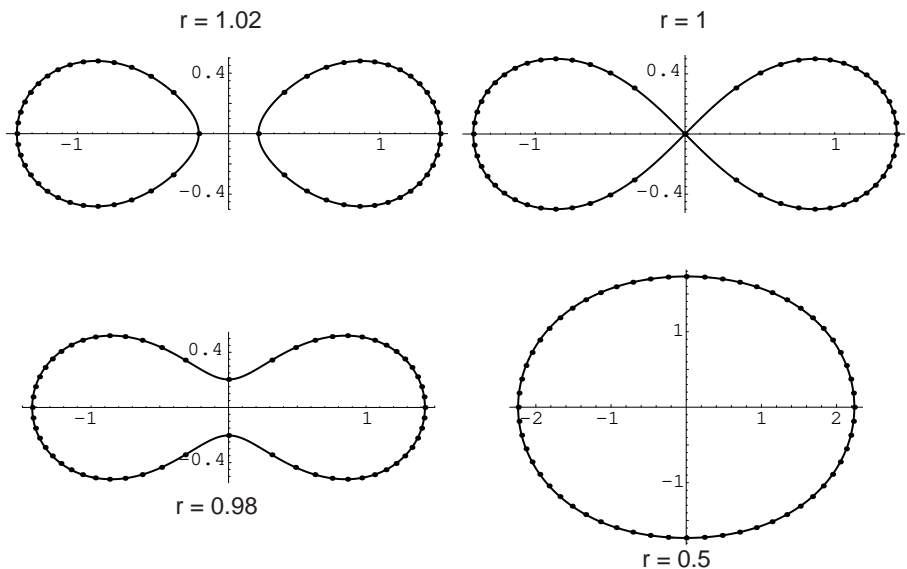,width=12cm}}
\caption{The enhan\c con for the distribution (\ref{newdis}) at $r=1.02$, 
$r=1$, $r=0.98$ and $r=0.5$. The dots correspond to the actual roots of the 
polynomial $p(x)^{2}-1/r^{2 N}$ for $N=30$. At $r=\rc =1$, two roots 
coincide, the two components of the enhan\c con merge, and we have a second 
class singularity. At small $r$ the enhan\c con is a circle of radius $1/r$.
\label{fig4}}
\end{figure}

If we take $r$ to be real for concreteness, then the critical value of $r$ 
is exactly given by
\begin{equation}
\label{rcritexa}
\rc =1
\end{equation}
and corresponds to the merging of the branching points
\begin{equation}
\label{exactroot}
x_{\pm} = \pm\sqrt{1-1/r^{2}}.
\end{equation}
The formulas (\ref{rcritexa}) and (\ref{exactroot}) are to be compared 
with (\ref{rcrit},\ref{rcritapp},\ref{xm},\ref{x1}). They exhibit one 
important difference between first class and second class singularities: 
the large $N$ expansion of the roots (or the shape of the enhan\c con) is 
perfectly smooth in the case of second class singularities. However, for 
the same general reasons as in the case of first class singularities, the 
large $N$ expansion does break down as well. To demonstrate this, we have 
to consider the Seiberg-Witten periods which are the
physical observables,
\begin{equation}
\label{swpnew}
z = {1\over i\pi}\, \int_{x_{-}}^{x_{+}}\! dx\, 
{xp'(x)\over\sqrt{p(x)^{2}-1/r^{2N}}} = -{N\over i\pi}\,
\int_{1/r^{2}}^{1} {du\over u}\, \sqrt{{1-1/(u r^{2})\over 1-u^{N}}}\cvp
\end{equation}
where $u=1/(r^{2} (1-x^{2}))$. A straightforward calculation along the 
lines of the preceding subsection or of the Appendix
then yields the large $N$ expansion
\begin{eqnarray}
\label{znewexp}
&&\hskip -1.75cm {i\pi z\over N} = 2\sqrt{1-1/r^{2}} - 2 \ln\left( 
1+\sqrt{1-1/r^{2}}\right) - 2\ln r - {2\sqrt{1-1/r^{2}}\,\ln 2\over N}
\nonumber\\
&&\hskip 6cm +{\pi^{2}/6 - 2 (\ln 2)^{2}\over 2 N^{2}r^{2}\sqrt{1-1/r^{2}}} +
{\cal O}\bigl(1/N^{3}\bigr).\\\nonumber
\end{eqnarray}
We get non-analytic terms at leading order, and genuine divergences at 
order $1/N^{2}$.

\section{Open problems}
We have seen that, despite the fact that the low energy effective action
of ${\cal N}=2$ super Yang-Mills is, in some sense, generated by
instantons only, it has a surprisingly rich and interesting large $N$
expansion. 

Low-dimensional sections of the moduli space can be easily
analysed. For example, Figure 2 is very reminiscent of the structure 
of the moduli space for $N=2$. This suggests that the spectrum of BPS
states and the curves of marginal stability
could have a simple description 
at large $N$, and that the methods of \cite{FB} may be useful.

We have elucidated the fate of large instantons at strong coupling and
large $N$ in ${\cal N}=2$ super Yang-Mills: they disintegrate into
fractional instantons. These fractional instantons are responsible for a
new class of contributions in the large $N$ expansion, generating a
series in $1/N$ in addition to the standard series in $1/N^2$ from
Feynman diagrams. These new contributions were brought to the fore
particularly clearly because we have studied observables 
which pick
up a single one-loop term from Feynman diagrams. More general observables
will probably have non-trivial contributions from both types of terms. 
Note that the leading corrections to $N=\infty$ are
dominated by fractional instantons. It would be very interesting
to find direct general arguments explaining why
fractional instantons contribute a well-behaved asymptotic
series in $1/N$. The fate of instantons in real world QCD remains of
course an open problem, but it is fascinating to contemplate the
possibility that phenomena qualitatively similar to the one studied in
the present paper could occur. This is
suggested by the fact that the basic difficulty with instantons---IR
problems due to the Landau pole for large instantons---are independent of
supersymmetry.

The standard 't Hooft analysis of perturbation theory at large $N$ in
$\suN$ gauge theories \cite{tHooft} is consistent with 
a dual description in
terms of closed oriented strings. Our analysis of the
fractional instanton contributions shows
that open strings must also be present. Though adding open strings to a
closed string theory is a non-trivial step, it 
is satisfying that fractional instanton terms
can be interpreted in a string picture at all.

The breakdown of the large $N$ expansion at singularities may come as a
surprise (it was anticipated
in the recent studies of two-dimensional models by
the author \cite{fer1,fer2,fer3}), but we have argued that 
the physics from the field theory point of view
is the same as the one that produces IR divergencies in
corrections to mean field theory below the critical dimension.  
On the other hand, the interpretation from the string theory point of
view is much less clear. A naive application of the UV/IR relation 
seems to indicate
that the string dual is not UV renormalizable, but this would 
be a rather strange property for a string theory. Another possibility, 
which was pointed out to me by I.~Klebanov, is that the divergencies are 
due to tensionless strings at the singularities. This interpretation 
requires the appearance of divergences (or at least non-analytic 
contributions) at leading order, which is not the case in (\ref{zfinal}). 
This might nevertheless be made consistent by correctly identifying the 
terms generated by closed strings or open strings diagrams. 
This point clearly deserves further investigation.

Another fascinating line of research, suggested
by the divergencies of the $1/N$
expansion near singularities, is that it might be possible to 
extract finite universal string 
amplitudes by taking $N\rightarrow\infty$ and 
$r\rightarrow\rc$ in a correlated way, along the lines of \cite{BK}.
A preliminary investigation will be presented
in a forthcoming paper \cite{fer4}.
\section*{Acknowledgements}
I would like to acknowledge useful discussions with Igor R.\ Klebanov and
Kostas Skenderis.
\begin{appendix}
\section*{Appendix: a simple toy integral}
Let us consider the integral
\begin{equation}
\label{In}
I_N = \int_0^1\! dx\, \Bigl({1\over\sqrt{1-x^N}}-1\Bigr)\cdotp
\end{equation}
By changing variable to $y=x^N$, $I_N$ is straightforwardly expressed in
terms of Euler beta function,
\begin{equation}
\label{Inex}
I_N = {1\over N}\, B(1/N,1/2) -1 = \sqrt{\pi}\,
{\Gamma(1+1/N)\over\Gamma(1/2+1/N)} -1\cdotp
\end{equation}
The large $N$ expansion can then be readily found,
\begin{equation}
\label{InN}
I_N = {2\ln 2\over N} + 
{2(\ln 2)^2 - \pi^2/6\over N^2} + {\cal O} \bigl(1/N^3\bigr). 
\end{equation}
The difficulty one encounters to understand the large $N$ asymptotic
expansion directly on the integral formula (\ref{In})
is the same as for the integral (\ref{zp})
studied in the main text: the integrand is exponentially small (an
instanton effect) except for a region of size $1/N$ near $x=1$. One then
use the variable $y$ defined by $x=1-y/N$, and expand the integrand,
which yields
\begin{equation}
\label{Inexpanded}
I_N= {1\over N}\,\int_0^N\! dy\,\Bigl( {1\over\sqrt{1-e^{-y}}} -1\Bigr)
-{1\over 4 N^2}\, \int_0^N\! dy\, {y^2 e^{-y}\over (1-e^{-y})^{3/2}}
+ {\cal O}\bigl( 1/N^3 \bigr).
\end{equation}
The integration region can be extended to $y\in [0,\infty[$ by neglecting
terms of order $e^{-N}$. The integral at order $1/N$ is elementary and
gives the $(2\ln 2)/N$ correction. The integral at order $1/N^2$ is of the
type of the integrals encountered in the main text. By going to the
variable $y'=1/\sqrt{1-e^{-y}}$ and integrating by part, it can be
related to the dilogarithm function ${\rm Li}_2$ and thus explicitly
evaluated.
\end{appendix}
\vfill\eject
\end{document}